# RECENT EXPERIMENTS AT NDCX-II: IRRADIATION OF MATERIALS USING SHORT, INTENSE ION BEAMS*

P. A. Seidl[†], Q. Ji, A. Persaud, E. Feinberg, B. Ludewigt, M. Silverman, A. Sulyman,
W. L. Waldron, T. Schenkel, Lawrence Berkeley National Laboratory, Berkeley, USA
J. J. Barnard, A. Friedman, D. P. Grote, Lawrence Livermore National Laboratory, Livermore, USA
E. P. Gilson, I. D. Kaganovich, A. Stepanov, Princeton Plasma Physics Laboratory, Princeton, USA
F. Treffert, M. Zimmer, TU Darmstadt, Darmstadt, Germany

*Abstract*

We present an overview of the performance of the Neutralized Drift Compression Experiment-II (NDCX-II) accelerator at Berkeley Lab, and summarize recent studies of material properties created with nanosecond and millimeter-scale ion beam pulses. The scientific topics being explored include the dynamics of ion induced damage in materials, materials synthesis far from equilibrium, warm dense matter and intense beam-plasma physics. We summarize the improved accelerator performance, diagnostics and results of beam-induced irradiation of thin samples of, e.g., tin and silicon. Bunches with over $3 \times 10^{10}$ ions, 1-mm radius, and 2-30 ns FWHM duration have been created. To achieve these short pulse durations and mm-scale focal spot radii, the 1.2 MeV He$^+$ ion beam is neutralized in a drift compression section which removes the space charge defocusing effect during final compression and focusing. Quantitative comparison of detailed particle-in-cell simulations with the experiment play an important role in optimizing accelerator performance; these keep pace with the accelerator repetition rate of ~1/minute.

## INTRODUCTION

Intense pulses of ions in the MeV range enable new studies of the properties of matter ranging from low intensity (negligible heating, but active collective effects due to proximate ion trajectories in time and space), to high intensity where the target may be heated to the few-eV range and beyond. By choosing the ion mass and kinetic energy to be near the Bragg peak, dE/dx is maximized and a thin target may be heated with high uniformity, thus enabling high-energy density physics (HEDP) experiments in the warm dense matter (WDM) regime. The Neutralized Drift Compression Experiment (NDCX-II) was designed with this motivation [1-3].

Reproducible ion pulses (N>10$^{11}$ /bunch), with bunch duration and spot size in the nanosecond and millimeter range, meet the requirements to explore the physics topics identified above. The formation of the bunches generally involves an accelerator beam with high perveance and low emittance, attractive for exploring basic beam physics of general interest, and relevant to the high-current, high-intensity ion beams needed for heavy-ion-driven inertial fusion energy [4].

Furthermore, short ion pulses at high intensity (but below melting) enable pump-probe experiments that explore the dynamics of radiation-induced defects in materials. For high peak currents and short ion pulses, the response of the material to radiation may enter a non-linear regime due to the overlapping collision cascades initiated by the incident ions. These effects may be transient (no memory effect at a subsequent pulse) and the short, intense pulses of ions provide an opportunity to observe the time-resolved multi-scale dynamics of radiation-induced defects [5-7]. In addition, by measuring the ion range during the course of the ion pulse, the effects of defects and heating on range can be observed. The time-resolved information provides insight and constraints on models of defect formation and in the design of structural materials, for example, for fission and fusion reactors.

## ACCELERATOR PERFORMANCE AND BEAM MODELING

A new multicusp, multiple-aperture plasma ion source is capable of generating high purity ion beams of, for example, protons, helium, neon and argon [8]. To date, we have used it solely for the generation of He$^+$ ions. Furthermore, helium at about 1 MeV is nearly ideal for highly uniform volumetric energy deposition, because particles enter thin targets slightly above the Bragg peak energy and exit below it, leading to energy loss in the target, uniform within several percent.

An ion induction accelerator is capable of simultaneously accelerating and rapidly compressing beam pulses by adjusting the slope and amplitude of the voltage waveforms in each gap. In NDCX-II, this is accomplished with 12 compression and acceleration waveforms driven with peak voltages ranging from 15 kV to 200 kV and durations of 0.07-1 µs [9]. The first seven acceleration cells are driven by spark-gap switched, lumped element circuits tuned to produce the required cell voltage waveforms. These waveforms ("compression" waveforms because of their characteristic triangular shape) have peak voltages ranging from 20 kV to 50 kV. An essential design objective of the compression pulsers is to compress the bunch to <70 ns so that it can be further accelerated and bunched by the 200-kV Blumlein pulsers which drive the last five acceleration cells.

In the final drift section, the bunch has a head-to-tail velocity ramp that further compresses the beam by an order-of-magnitude. The space-charge forces are sufficiently high at this stage to require that dense plasma, generated prior to the passage of the beam pulse, neutralizes the beam self-field and enables focusing and bunching of the beam to the millimeter and nanosecond range [10]. The beam perveance:

$$Q = \frac{2qI}{4\pi\varepsilon_o M v^3}$$

is high throughout the accelerator ($\approx 2\times10^{-3}$) at the first acceleration gap and higher as the beam compresses. When passing through the neutralizing plasma the coasting beam compresses with negligible space-charge repulsion, the spot size and duration are limited mainly by voltage waveform fidelity and the chromatic aberrations and emittance of the beam.

___________

* PASeidl@lbl.gov Work supported by the US DOE under contracts DE-AC0205CH11231 (LBNL), DE-AC52- 07NA27344 (LLNL) and DE-AC02-09CH11466 (PPPL).

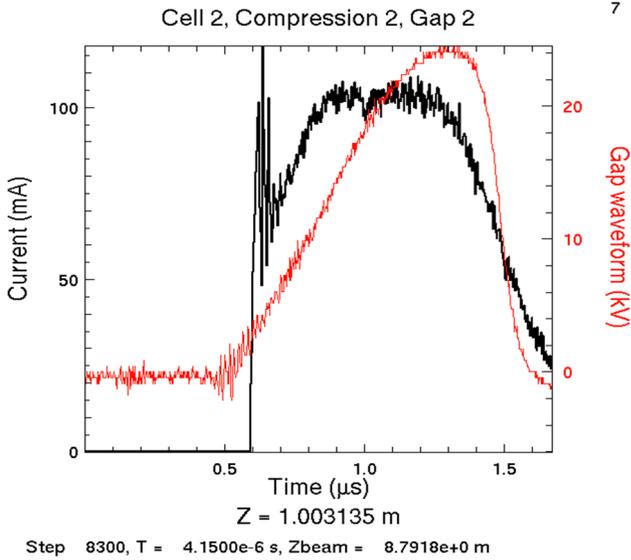

Figure 1: Warp PIC simulations are initialized with measured acceleration gap waveforms as shown in the example above.

The accelerator provides a novel platform to extend the limits of intense beam and beam-plasma physics. For example, the ion-electron two-stream instability has been predicted [11], and it may be observed in NDCX-II by passing a nearly constant energy beam through a plasma. The manifestation would be a significant transverse defocusing and longitudinal bunching of the specially prepared beam. In NDCX-II, the effect is normally absent because of the imposed velocity ramp on the beam distribution.

Another interesting opportunity would be to demonstrate the collective focusing of an ion beam in a weak magnetic field [12]. The focusing occurs due to the formation of a strong radial self-electric field due to the rearrangement of the plasma electrons moving with the beam, and in response to a weak magnetic field. The magnetic field is established by the final solenoid near the end of the neutralized drift section. But, instead of using the full strength of the final solenoid (5-8 Tesla) to focus the beam, equivalent focusing may be achieve with only ~0.02 Tesla. If demonstrated to be practical, the focusing magnet strength would be greatly reduced with a corresponding reduction of the stray magnetic field at the nearby target plane – as preferred for some beam-target interaction experiments.

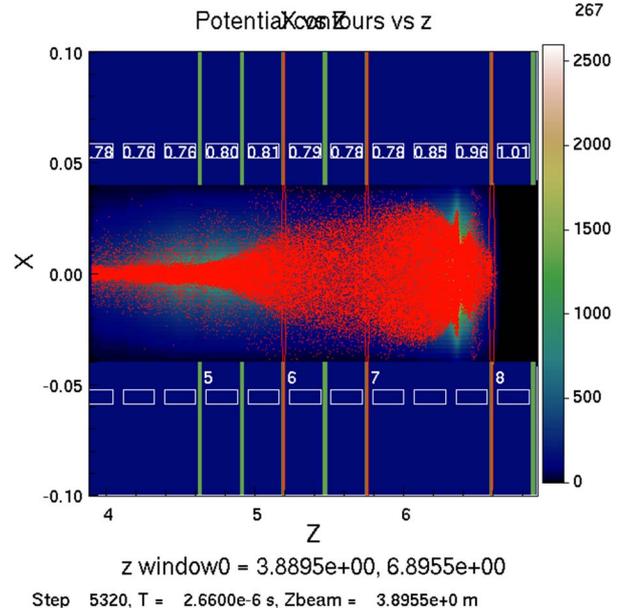

Figure 2: The simulations are initialized with measured magnetic field strengths. The x-z projections of the beam distribution show the main features of the particle transverse distribution as well as halo particle loss. The red contour lines show the equipotentials from the applied acceleration fields. The intensity bar (right) indicates the space charge potential.

Also, double pulses of $K^+$ or $Li^+$ ions, separated by 0.05 - 0.5 μs, have been created from single triggers of the high voltage pulses in the injector and accelerator [6]. Single injected bunches are cleaved due to the shifted timings of the firing of the compression waveforms. Double pulses are of interest for example, in pump-probe experiments of radiation effects in matter.

We have enhanced the integration of simulations with the experimental data acquisition via automatic import from the data acquisition database of injector and acceleration waveforms from the experiment, along with all of the particular focusing magnet settings. Thus the influence of small imperfections (timing, waveform shape) in the acceleration waveforms is included in the simulations which in turn provides more accurate predictions of the accelerator performance. Figure 1 shows the measured acceleration voltage for a particular shot applied to the acceleration gap. The Warp simulation [1] also shows the arrival time of the beam pulse at the same gap – valuable for evaluating the effects of pulser timing fine tuning. In Fig. 2 the particle distribution is shown for the same pulse through several solenoids downstream of Fig. 1. We have demonstrated that the simulations can keep up with the accelerator repetition rate of $\approx$1/min with parallel runs running on a handful of processors.

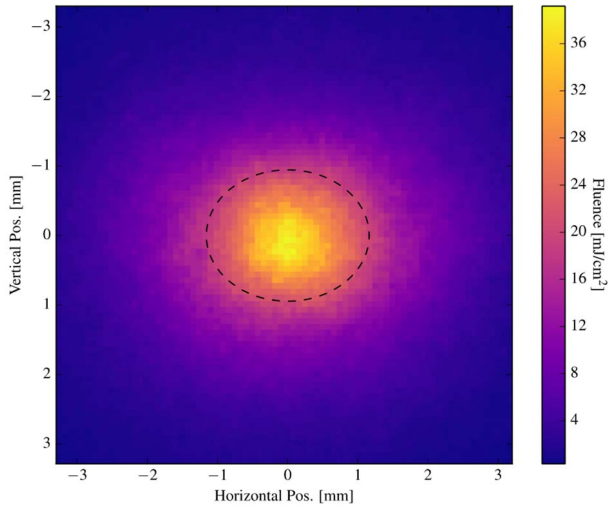

Figure 3: Beam intensity imaged by an intensified CCD camera. The beam is intercepted by an Al2O3 scintillator at the target plane. The dotted line shows the 2•rms radius.

Figure 3 shows the beam current density profile measured with a scintillator and gated CCD camera for a pulse with $2.5 \times 10^{10}$ ions, 6 ns FWHM and and 2•rms beam radius is ≈1 mm.

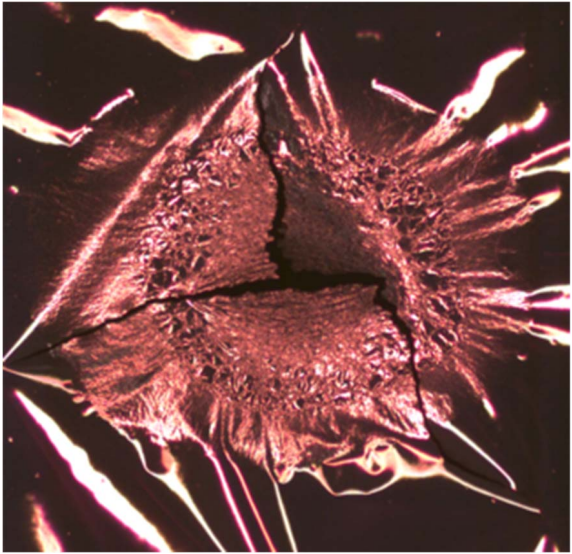

Figure 4: A microscope image of the Sn foil after being struck by a helium beam pulse shows amorphization after rapid heating and subsequent cooling near the onset of the melting point. The field of view is ≈ 6mm.

The envelope and beam current vary significantly through the pulse; the solenoid focusing is mostly balancing the transverse space charge. The magnetic fields are chosen to minimize envelope mismatch and transmit the highest charge to the target plane with the best focal properties. Simulations suggest that the total bunch charge can be increased several fold with modest adjustments to the solenoid and compression voltage waveforms.

## TARGET EXPERIMENTS

Figure 4 shows an example of target heating on a nanosecond time scale. The 0.3-μm tin target (Fig. 4) after one shot with charactaristics of that in Fig. 3 shows evidence of intense heating and possible amorphization at high ion fluences. The peak fluence is predicted to heat thin samples uniformly through by several hundred degrees, just below the onset of melting.

With the range of intensities and dose rates presently available, there is an opportunity to create novel states of matter with the NDCX-II beam, such as local nitrogen-vacancy center formation in nitrogen-implanted diamond by transient heating, followed by rapid quenching [13]. Annealing of specially prepared materials often takes place on a timescale of seconds to minutes, such as in the formation of paramagnetic materials [14]. With NDCX-II, we have an opportunity to locally excite materials with intense ion pulses on a ns timescale and stabilize novel phases by rapid quenching.

## REFERENCES

[1] A. Friedman, et al. "Beam dynamics of the Neutralized Drift Compression Experiment-II, a novel pulse-compressing ion accelerator," *Phys. Plasmas* 17 056704 2010.

[2] J. J. Barnard, et al. "NDCX-II target experiments and simulations," *Nucl. Inst. Meth.* A733 45 2014.

[3] W. L. Waldron, et al. "The NDCX-II engineering design," *Nucl. Inst. Meth*. A733 2014.

[4] R. O. Bangerter, A. Faltens and P. A. Seidl, "Accelerators for Inertial Fusion Energy Production," *Rev. Accel. Sci. & Tech*. 6, 85 2013.

[5] T. Schenkel, *et al*., "Towards pump-probe experiments of defect dynamics with short ion beam pulses," *Nucl. Inst. Meth.* B315 350 2013.

[6] A. Persaud, *et al.*, "Accessing Defect Dynamics using Intense, Nanosecond Pulsed Ion Beams," *Phys. Procedia* 66 604 2015.

[7] X. M. Bai, *et al*. "Efficient Annealing of Radiation Damage Near Grain Boundaries via Interstitial Emission," *Science* 327, 1631 2010.

[8] Q. Ji, *et al*, "Development and testing of a pulsed helium ion source for probing materials and warm dense matter studies," *RSI* 87 (2) 02B707 2016.

[9] P. A. Seidl, *et al*., "Short intense ion pulses for materials and warm dense matter research," *NIM* A800 2015

[10] E. P. Gilson, *et al. Nucl. Inst. Meth*. A733 75 2013.

[11] E. Tokluoglu and I. D. Kaganovich, "Defocusing of an ion beam propagating in background plasma due to two-stream instability," *Phys. Plasmas* 22 040701 2015.

[12] M. Dorf, *et al*., "Enhanced Collective Focusing of Intense Neutralized Ion Beam Pulses in the Presence of Weak Solenoidal Magnetic Fields," *Phys. Plasmas* 19, 056704 2012.

[13] J. Schwartz, *et al.* "Local formation of nitrogen-vacancy centers in diamond by swift heavy ions," *J. Appl. Phys*. 116, 214107 2014.

[14] A. Bienfait*, et al*. "Reaching the quantum limit of sensitivity in electron spin resonance," *Nature Nano*. 11, 253–257 2016.